\input harvmac

\overfullrule=0pt
\parindent=0pt

%%%%%%%%%

\def\RR{{$R\otimes R$}}

\def\det{\hbox{\rm det}}

\def\G(#1){\Gamma(#1)}

\def\C|#1{{\cal #1}}
\def\(#1#2){(\zeta_#1\cdot\zeta_#2)}

%%%%% journals %%%%

\def\xxx#1 {{hep-th/#1}}
\def\lr { \lref}
\def\npb#1(#2)#3 { Nucl. Phys. {\bf B#1} (#2) #3 }
\def\rep#1(#2)#3 { Phys. Rept.{\bf #1} (#2) #3 }
\def\plb#1(#2)#3{Phys. Lett. {\bf #1B} (#2) #3}
\def\prl#1(#2)#3{Phys. Rev. Lett.{\bf #1} (#2) #3}
\def\physrev#1(#2)#3{Phys. Rev. {\bf D#1} (#2) #3}
\def\ap#1(#2)#3{Ann. Phys. {\bf #1} (#2) #3}
\def\rmp#1(#2)#3{Rev. Mod. Phys. {\bf #1} (#2) #3}
\def\cmp#1(#2)#3{Comm. Math. Phys. {\bf #1} (#2) #3}
\def\mpl#1(#2)#3{Mod. Phys. Lett. {\bf #1} (#2) #3}
\def\ijmp#1(#2)#3{Int. J. Mod. Phys. {\bf A#1} (#2) #3}

\def\tr{{\rm tr}}

\def\Rfour{t_8t_8R^4}
\def\lam16{\lambda^{16}}
\def\psih{\hat \psi}

\parindent 25pt
\overfullrule=0pt
\tolerance=10000

\sequentialequations

\lr\greenschwarz{M.B.~Green and  J.H.~Schwarz, {\it  
Supersymmetrical String
Theories}, \plb109(1982)444.}
\lr\seibergwit{N.~Seiberg and E.~Witten, {\it Electric - magnetic  
duality,
monopole condensation, and
                  confinement in N=2 supersymmetric Yang-Mills theory},
\npb426(1994)19, \xxx9407087; $\;$ {\it Monopoles, duality and  
chiral symmetry
breaking in N=2 supersymmetric QCD}, \npb431(1994)484, \xxx9408099.}

\lr\feretti{ U.H.~Danielsson, G.~Ferretti and  B.~Sundborg, {\it
D-particle
Dynamics and Bound States},  Int.J.Mod.Phys. {\bf A11} (1996) 5463,
\xxx9603081.}
\lr\pouliot{D. Kabat and P. Pouliot, {\it A Comment on Zero-brane Quantum
Mechanic}, \prl77(1996)1004, \xxx9603127. }
\lr\matrixth{T.~Banks, W.~Fischler, S.H.~Shenker and  L.~Susskind,
{\it M
Theory As A Matrix Model: A Conjecture}, \physrev55(1997)5112,
\xxx9610043. }
\lr\nekrasov{N.~Nekrasov and A.~Lawrence, {\it  Instanton sums and five
dimensional gauge theories
}, \xxx9706025.}
\lr\dorey{ N.~Dorey, V.V.~Khoze and  M.P.~Mattis, {\it
Supersymmetry and the
Multi-Instanton Measure}, \xxx9708036. }
\lr\wittb{E.~Witten, {\it String theory dynamics in various dimensions},
\npb443(1995)85, \xxx9503124.}
\lr\porrati{M.~Porrati and A.~Rozenberg, {\it Bound states at
threshold in
supersymmetric quantum  mechanics}, \xxx9708119.}
\lr\gg{M.B.~Green and M.~Gutperle, {\it Effects of D-instantons},
\npb498(1997)195,
\xxx9701093.}
\lr\ggII{M.B.~Green and M.~Gutperle, {\it Configurations of two
D-instantons},
 \plb398(1997)69, \xxx9612127.}
\lr\ggv{M.B.~Green,  M.~Gutperle and P.~Vanhove, {\it One loop in
eleven-dimensions}, \plb409(1997)177, \xxx9706175.}
\lr\gv{M.B.~Green and P.~Vanhove, {\it D-Instantons, strings and M
theory},
\plb408(1997)122, \xxx9704145}
\lr\ggk{M.B.~Green,  M.~Gutperle and H.~Kwon, {\it Sixteen fermion
and related
terms in M theory on $T^2$}, \xxx9710151. }
\lr\witt{E.~Witten, {\it Bound States Of Strings And $p$-Branes},
\xxx9510135,
\npb460(1996)335.}
\lr\ggp{G.W.~Gibbons, M.B.~Green and M.J.~Perry, {\it Instantons and
seven-branes in type IIB superstring theory}, \plb370(1996)33,
\xxx9511080 }
\lr\yia{P. Yi,{\it Witten index and threshold bound states of
D-branes},
\xxx9704098 }.
\lr\sethia{S.~Sethi and M.~Stern, {\it D-brane bound states redux},
\xxx9705046.
}
\lr\gw{D.J.~Gross and E.~Witten, {\it Superstring modifications of
Einstein's
    equations}, \npb277(1986)1.}
\lr\unit{N.~Sakai and Y.~Tanii, {\it One-loop amplitudes and
effective action
in
superstring theories}, \npb287(1987)457.}
\lr\gris{M.T.~Grisaru , A.E.M~Van de Ven and D.~Zanon, {\it
Two-dimensional
supersymmetric sigma models on Ricci flat Kahler manifolds are not
finite},
\npb277(1986)388 ; {\it Four loop divergences for the N=1 supersymmetric
nonlinear sigma model in two-dimensions}, \npb277(1986)409.}
\lr\antoniadis{I. Antoniadis, B. Pioline and T.R. Taylor, {\it Calculable
$e^{1/\lambda}$ effects}, \xxx9707222.}
\lr\berkov{N. Berkovits, {\it Construction of $R^4$ terms in N=2 D=8
superspace}, \xxx9709116.}
\lr\bachas{ C.~Bachas, {\it Heterotic versus Type~I\/}, hep-th/9710102;\
C.~Bachas and P.~Vanhove, {\it The Rules of D-brane Calculus\/},
    CPTH-S587.1297, to appear.}
%%%%%%%%%%%%%%%%%%%
\noblackbox
\baselineskip 14pt plus 2pt minus 2pt
\Title{\vbox{\baselineskip12pt
\hbox{hep-th/9711107}
\hbox{DAMTP-97-121}
\hbox{PUPT-1745}
}}
{\vbox{
\centerline{D-particle bound states and }
\vskip 0.2cm
\centerline{the   D-instanton measure} }}
\centerline{ Michael B. Green,}
\medskip
\centerline{DAMTP, Silver Street, Cambridge CB3 9EW, UK} \centerline{\it
M.B.Green@damtp.cam.ac.uk} \bigskip
\centerline{Michael Gutperle}
\medskip
\centerline{Joseph Henry Laboratories, Princeton University}
\centerline{Princeton, New Jersey 08544,  USA}
 \centerline{\it gutperle@feynman.princeton.edu}

\bigskip
 \medskip

\centerline{{\bf Abstract}}

A connection is made between the  Witten index
of  relevance  to threshold bound states of
D-particles in
the type IIA superstring theory and the  measure that enters
D-instanton sums for processes dominated by single multiply-charged
D-instantons in  the  type IIB theory.

%%%%%%%%%%%%%%%%%%%%%%%%%%%%%%%%%%%%%%%%%%%%%%%%%%%%%%%%%%%%%%%%%%%
\noblackbox
\baselineskip 14pt plus 2pt minus 2pt

\Date{November 1997}

\newsec{Introduction}

Since D-particles in ten dimensions are Kaluza--Klein modes of
eleven-dimensional supergravity there must be precisely  one
normalizable
D-particle state with charge $N$
\wittb.  This strongly suggests  that there
should be one
threshold bound state of charge $N$ and mass $Ne^{-\phi}$ in the
supersymmetric $SU(N)$
Yang--Mills quantum mechanical system that  describes the low
energy
dynamics of  $N$ interacting
singly-charged (\lq fundamental')  D-particles \feretti,\pouliot.
The existence
of these bound
states is an essential link in the interconnecting duality chains
that relate
all the string theories and eleven-dimensional supergravity and it
is one of
the
key ingredients in the matrix model \matrixth.
Up to now it has
been shown
\yia,\sethia\ that a suitably defined  Witten index, $\tr (-1)^F$,
is equal to
one in the two fundamental  D-particle system.  This  implies that
there is at
least one threshold bound state in the two D-particle system.
Further evidence
for   the
existence of  a bound state  when  $N$ is prime was given in \porrati.

The world-line of a charge-$n$  D-particle of the euclidean type
IIA string
theory winding $\hat m$ times
around  a compact dimension may be identified, via T-duality,  with a
D-instanton
of the compactified IIB theory \gg.  The euclidean D-particle
action translates
into an expression for the  action of  a
D-instanton  with
instanton number $N = \hat m n $ of the form
\eqn\instactt{S^{N} = 2\pi i \rho \hat m n,}
where   $\rho = C^{(0)} + i e^{-\phi^B}$ is the complex scalar
field of type
IIB supergravity  ($\phi^B$ is the IIB dilaton
and $C^{(0)}$
is the Ramond--Ramond (\RR) scalar).   The spectrum of
D-particle   states is
thereby related directly to the measure  of multiply-charged
D-instantons.
It is therefore natural to ask if information concerning the moduli
space of
D-instantons might prove useful in understanding the D-particle
bound states.

Such information is contained in certain terms in the M-theory or type II
string theory effective actions that are plausibly given entirely by
tree-level, one-loop and non-perturbative D-instanton corrections
\gg,\gv,\ggv.
These terms include a  $f(\rho,\bar \rho)R^4$ term  (where $R^4$
indicates a
particular contraction of four Riemann curvatures), a $f_{16}
(\rho,\bar\rho)
\lam16$ term  \ggk\ (where $\lambda$ is the spin-1/2 fermion in the
type IIB
theory)  and other terms related to these by supersymmetry. The functions
$f(\rho,\bar\rho)$ and $f_{16} (\rho,\bar\rho)$ are non-holomorphic
modular
forms  of appropriate weight.   The sum over the infinite set of
D-instantons  which emerges from the asymptotic behaviour of $f$
and $f_{16}$
includes a measure factor that  should be related to the
calculation of the
Witten index of the D-particle bound state problem.  This
connection is the
subject of this paper.

\newsec{ D-instanton moduli space}

Classical D-instantons  are   solutions of  euclidean  IIB
supergravity
\ggp\  which  preserve  half the euclidean supersymmetry in which all the
fields
are trivial (in Einstein frame)  apart form the
complex scalar.   The  solution describing $k$ D-instantons located at
space-time
points $y^\mu_i$ ($\mu =0, \cdots, 9$ and $i=1, \cdots, k$) each  
carrying an
integer `charge' $N_i$   is defined by the dilaton profile,
\eqn\multicenter{e^{\phi(x)-\phi_{\infty}}= 1+ \sum_{i=1}^k
{2N_i e^{-\phi_\infty}\over \pi^{3/2} |x-y_i|^8}}
together with  $dC^{(0)} = - id e^{-\phi}$.
The  string coupling constant is identified  with the asymptotic  
value of the
dilaton,  $g =
e^{\phi_{\infty}}$   and the
action for this configuration is given by
 $S^{\{N_i\}} = \sum_i 2\pi i  N_i \rho_0$,  where  $\rho_0 = \chi + i
/g$ is the
asymptotic value of $\rho$.

Sixteen of the   thirty-two  IIB
supersymmetries are conserved in the background of a
single  charge-$N$  D-instanton while the   broken supersymmetries  
generate
fermionic  collective
coordinates   which are the super-partners of the
bosonic collective coordinates, $y^\mu$.  The integration over  
these fermionic
zero modes induces new  local interactions (as with t'Hooft
vertices).     Multi-instanton solutions  have more bosonic and
fermionic zero
modes and  therefore  contribute to effective interaction
vertices with
more
derivatives.   In the following we will be considering interaction  
terms  that
are  integrals over sixteen  fermionic coordinates, or half the  
superspace,
and
are
therefore
dominated by the single D-instantons.
These interactions  are
expected to be protected by non-renormalization theorems.

As is well known, the classical supergravity solutions describing  
D-branes  are
only appropriate for
describing  physics at large separations.  At small D-brane  
separations   the
appropriate dynamical degrees of freedom are the open strings
connecting them and  the low energy excitations of
the $N$ D-brane  system are
associated with the fermionic and bosonic open-string ground  
states.    The
D-brane effective world-volume
theory  is thereby identified with the dimensional reduction of  
ten-dimensional
$U(N)$  supersymmetric
Yang-Mills   theory to $p+1$ dimensions \witt.   The situation with
D-instantons is somewhat degenerate since $p=-1$
and  the world-volume reduces to a space-time point.  In this case  
the open
strings have end-points fixed in all space-time dimensions so they do not
describe propagating degrees of freedom.
The configuration space of $N$ D-instantons is determined by bosonic and
fermionic $U(N)$ matrices $A_\mu$ and $\psi_\alpha$, where $\mu$
and $\alpha$
denote  $SO(10)$
vector and spinor indices, respectively. The
reduction of the supersymmetric Yang--Mills action to a point is given by
\eqn\DinstSYM{S=  {1\over 4}
\tr([A_\mu,A_\nu]^2)+{i\over
2}\tr(\bar{\psi} \Gamma^\mu [A_\mu,\psi]).}

The `center of mass' degrees of freedom  are
associated with
the element of the Cartan subalgebra  of $U(N)$ proportional to the
unit matrix  and do not appear
in the action \DinstSYM.  They do, however,  enter as shift  
symmetries in the
supersymmetry transformations,
\eqn\susytrafo{\eqalign{& \delta A_\mu = i \bar{\eta}\Gamma_\mu \psi,\cr
 & \delta \psi = [A_\mu,A_\nu]\Gamma^{\mu\nu}\eta + {\bf 1}
\,\epsilon.  }}
The sixteen-dimensional spinor $\eta$ parameterizes the dimensionally
reduced
supersymmetry of the $SU(N)$ YM theory and the spinor $\epsilon$
acts as a
constant
shift on the fermions in the `center of mass' degrees of freedom.
Hence the
U(1)
part of $U(N)$ fermionic fields $\psi$ plays the role of the
sixteen  fermionic
collective coordinates associated with the charge-$N$ D-instanton.

After factoring out the centre of mass coordinates the $N$ D-instanton
partition function
 is given by
\eqn\integSYM{Z = \int d^{10}y \int
d^{16}\epsilon\,  Z_{SU(N)},}
where
\eqn\zdef{Z_{SU(N)} =  \int_{SU(N)} D\psi DA \exp(-S_{SYM}[A,\psi]).}
The integration over $y^\mu$ and $\epsilon^A$  is the integral
over the overall bosonic and fermionic collective coordinates of  the
collection of D-instantons  of total  charge $N$.   A non-zero value for
$Z_{SU(N)}$ can only arise from configurations in which there are
no extra fermionic
zero modes (in addition to  $\epsilon$) which is characteristic of
a single
multiply-charged D-instanton.   The quantity $Z_{SU(N)}$ should  
therefore be
identified with the measure on the space of single charge-$N$  
D-instantons.  As
we shall see, the integral \zdef\
is the same  as the
bulk integral which appears in the calculation of the Witten index
in the bound  state problem of D-particles. Although this integral  
has not been
evaluated for general $N$ it has been  explicitly evaluated in the   
case  $N=2$
    \yia,\sethia.  The sub-integral over  configurations in which  
the elements
of the Cartan subalgebra, $A_\mu^3, \psi^3$, are fixed  was  
evaluated in \ggII\
and  corresponds to  two D-instantons at a fixed separation.

\newsec{D-instanton terms in the effective action}

The presence of a D-instanton induced $R^4$  vertex in IIB string  
theory  was
demonstrated in \gg.     This, together with the known   
perturbative string
tree-level  \gris,\gw\  and
one-loop \greenschwarz,\unit\  $R^4$ terms,  motivated the conjecture  
that the
exact
$R^4$ term in IIB superstring theory is
given by
\gg\
\eqn\rfour{S_{R^4}= \int d^{10} x \sqrt{g} \rho_2^{1/2}  
f(\rho,\bar{\rho})
t_8t_8 R^4
,}
where $g_{\mu\nu}^B$ is the ten-dimensional IIB string-frame metric  
and  the
function
$f$ is a non-holomorphic modular function defined by
\eqn\maas{ f(\rho,\bar{\rho})=\sum_{(n,m)\neq(0,0)}{\rho_2^{3/2}\over
|m+n\rho|^3}.}
The large-$\rho_2$ (small coupling)  expansion of $f$ is given by
\eqn\newsum{\eqalign{ & \rho_2^{1/2} f (\rho, \bar \rho) =
2\zeta(3)(\rho_2)^{2} +
{2\pi^2\over
3}  + 4\pi^{3/2} \sum_{N}(N\rho_2)^{1/2}\sum_{N|\hat{m}}{1\over  
{\hat{m}}^2}
\cr
&
\times  \left(e^{2\pi i N
\rho} + e^{-2\pi i N \bar \rho} \right) \left(1 + \sum_{k=1}^\infty
(4\pi N
\rho_2)^{-k} {\Gamma( k -1/2)\over \Gamma(- k -1/2) k!} \right) ,\cr}}
where $\sum_{N|n}$ denotes the sum over divisors of $N$.
This expression reproduces precisely the
 perturbative tree-level and one-loop
terms  and   suggests a  perturbative
non-renormalization theorem   beyond one loop for  $R^4$ and
other  terms
related by supersymmetry \gg.  Recently,  other arguments have lent  
 support to
this
\antoniadis,\berkov.
The non-perturbative terms  in \newsum\ are contributions of
D-instantons
(and
anti-D-instantons) of arbitrary charge $N$ together with an infinite
sequence  of
perturbative fluctuations around
each instanton configuration.  The expression obtained in \gg\ is
reproduced by writing  $N = n \hat m$,
for integer $\hat m$.   Following T-duality on an euclidean circle  
$\hat m$ can
be reinterpreted as the winding number of a charge-$n$ D-particle    
world-line
as described in the introduction.

Similar remarks apply to  other terms related by supersymmetry to  
the $\Rfour$
term, such as the sixteen-fermion
$\lam16$ term
(where $\lambda$ is the complex chiral spin-1/2 fermion of the IIB  
theory).
Since $\lambda$  carries charge $3/2$  under the $U(1)$ R-symmetry  
of the IIB
theory  it transforms under $SL(2,Z)$ with holomorphic and  
anti-holomorphic
weights $(3/4,-3/4)$ so  that  $\lambda^{16}$  has weight $(12,-12)$.  An
argument was given in \ggk\ that the  $\lam16$ contribution to the  
IIB action
has the  form
\eqn\lamsixteen{\int d^{10} x \sqrt{g} \rho_2^{1/2} \hat K f_{16}  
(\rho,\bar
\rho) \lambda^{16} ,}
where $\hat K$ is a kinematic coefficient  and $f_{16}$ is a specific
non-holomorphic modular form that  transforms  with holomorphic and
anti-holomorphic weights $(-12,12)$ under $SL(2,Z)$.     More generally,
supersymmetry relates a large number of terms of the same dimension  
which are
characterized by their $U(1)$ charge, which is an even integer, $2r$.
The   functions of $\rho$ and $\bar{\rho}$
 which multiply these terms  in the action are generalized   
non-holomorphic
modular forms, $F_{(r,-r)}(\rho,\bar{\rho})$,  with  weights $(r,-r)$.
These are  related to $f(\rho,\bar{\rho})$    \maas\ by
\eqn\covdera{F_{(r,-r)}(\rho, \bar{\rho})= \rho_2^r D^r
f(\rho,\bar{\rho}),}
where $D$ is the (nonholomorphic) covariant derivative that maps  
generalized
modular forms of weight $(l,l^\prime)$ to forms of weight  
$(l+2,l^\prime)$. The
action of $D$ on the generalized holomorphic forms $F_{(l,-l)}$ is  
defined by
 \eqn\covderb{D F_{(l,-l)}\equiv i\left({\partial\over \partial  
\rho}+ {l\over
\rho-\bar{\rho}}\right)F_{(l,-l)}.}
The modular form that enters into the  $\lambda^{16}$ interaction
is given by  $f_{\lambda^{16}}=F_{(12,-12)}(\rho,\bar{\rho})$ \ggk.  The
leading
$N$-instanton contributions to the large-$\rho_2$ expansion of the scalar
function
$F_{(r,-r)}$  can easily be determined from
\newsum\ and \covdera\ to be given by
\eqn\newsumf{ F_{(r,-r)}  \sim C_k \sum_N(N\rho_2)^{r+1/2}e^{2\pi i
N\rho}\sum_{N|\hat{m}}{1\over {\hat{m}}^2} +\cdots,}
where $C_k$ is a  constant and  $\cdots$ represents  perturbative
corrections to the instanton contribution that  are  suppressed by  
powers of
the
inverse instanton action, $1/(N\rho_2)$.
We wish to focus on the  measure factor,
\eqn\common{\sum_{N|\hat{m}} {1\over {\hat{m}}^2},}
which is common to all processes (all $k$) and should therefore be  
identified,
up to an
overall constant,  with the measure of singly-charged D-instantons,
$Z_{SU(N)}$.

\newsec{D-particles and D-instantons}
  The Witten index for the system of $N$ D-particles
was defined in \yia\sethia\ by
\eqn\indexa{\eqalign{ I^{(N)}&= \int dx \lim_{\beta\to \infty} \tr (-1)^F
e^{-\beta
H}(x,x)\cr
&= \lim_{R\to \infty} \lim_{\beta\to 0} \left\{
\int_{|x|<R}dx\;\tr(-1)^F e^{-\beta
H}+ {1\over2} \int_{|x|=R}dx \int_\beta^\infty d\beta^\prime\; \tr  
e_n (-1)^F Q
e^{-\beta^\prime
H}\right\},}}
where the first term defines a bulk contribution $I^{(N)}_{bulk}$ (which
in general
is not an
integer) and the second term a deficit contribution $I^{(N)}_{def}$  
(which
corrects the bulk term).   The separation  of the integer  $I^{(N)}$
into these two
parts is dependent on the order of limits.  We shall always consider the
infinite volume limit to be taken before the limit $\beta \to 0$.

The trace in \indexa\ is defined over gauge invariant states of the
Hilbert
space, which may be written as a trace over all  states  if a
projector
onto gauge
invariant states is inserted.  For the bulk term this gives
\eqn\fulltrace{ I^{(N)}_{bulk}= \lim_{\beta\to 0}\int_{SU(N)} d\eta \int
dx \; \tr
(-1)^F
e^{i \eta C}
e^{-\beta H}(x,x).}
Using the heat kernel approximation for the propagator in the
hamiltonian  $H$,
which is valid in
the limit $\beta\to 0$, it was observed in \yia\ and \sethia\ that
the exponent
of
the bulk part
of the index can be written in a $SO(10)$ invariant form. This is
possible
because
the
parameter $\eta$ turns into a tenth bosonic coordinate $A_0$.  The
resulting
integrals are  by closer inspection  exactly equivalent
 to the $SU(N)$ integral of the D-instanton collective coordinates  in
$Z_{SU(N)}$ \zdef.  This is simply the statement of T-duality for
the bulk
term,
since in \fulltrace\ the radius $\beta$ of the euclidean circle
shrinks to zero
size. Note that the insertion of $(-1)^F$ enforces supersymmetric
boundary
conditions on the fermionic fields.

The conjectured form for the measure, $Z_{SU(N)}$ given by \common,  
therefore
suggests that
the value of the bulk integral is given, for arbitrary $N$, by
\eqn\result{I^{(N)}_{bulk}= \sum_{N|\hat{m}}{1\over {\hat{m}}^2},}
where an overall constant has been fixed by noting that  
$I_{bulk}^{(1)} = 1$.
Note that for $N=2$ the value of  $I^{(2)}_{bulk}=5/4$ in \result\  
agrees with
the explicit
calculation of \yia,\sethia. When $N$ is prime the conjectured
 expression for  the bulk integral is
\eqn\bulkprime{I^{(N)}_{bulk}=\left( 1+{1\over N^2}\right).}

\newsec{The Deficit  Term}
The appearance  of  a deficit
term is related to the presence of boundary terms produced by an
integration by parts
that is necessary  in moving  $Q$ around in the
trace in \indexa. The only sizeable
contribution to the  boundary term  comes
from propagation along the flat directions of the potential.  A heuristic
evaluation  of the deficit
term in the $SU(2)$ case was given  by
Yi \yia\ who argued  that it  comes from the region in which the two
D-particles
are separated and the $SU(2)$ is broken to $U(1)$.  He suggested   
that the
particles might  behave as identical free particles in this region,  with
moduli
space $R^9/Z_2$.  Since the Witten index vanishes for free particles the
deficit
term, $I^{(2)}_{def}$, must cancel the  bulk  term for the free system,
$I^{(2)}_{0\, bulk}$.  But this  free  bulk term is easily  
evaluated and is
equal to $1/4$.  A more precise discussion  of the two D-particle  
system  based
on an analysis of the integration over the massive modes proves that the
deficit
term is indeed determined by free propagation on $R^9/Z_2$
 \sethia.     In the following we
shall assume that this prescription generalizes to  $N>2$   
D-particles so that,
for {\it prime} values of $N$,
\eqn\ndeff{ I_{def}^{(N)} = -  I_{0\, bulk}^{(N)},}
where $I_{0\, bulk}^{(N)}$ is the bulk index for $N$ identical free  
particles
moving on $R^{9(N-1)}/S_N$.   For {\it non-prime} values, $N=\hat m  
n$,  the
generalization will take into account the regions of moduli space  
of $\hat m$
free charge-$n$ particles on $R^{9(\hat m-1)}/S_{\hat m}$.

In the $N=2$ case considered in \yia,\sethia\ the only configuration that
contributed to the trace over the free two-particle states   was  
one with an
odd
permutation of the two particles. If the trace is expressed as a  
functional
integral this includes only  configurations in which the two  
D-particles  are
described by a single  euclidean world-line that winds twice around  
the compact
$\beta$ direction.     In order to generalize this to arbitrary $N$  
we need to
consider the action of $S_{N}$, which is the
 Weyl group of $SU(N)$.  This can be
parameterized
by matrices $M_{ab}$  acting  on the positions
of the $N$ particles  modded out by the overall translation
invariance, $X^i_a$ ($i=1, \cdots, 9$),  and the
fermions $\psi^\alpha_a$. The index $a=1,..,N-1$ labels the different
$U(1)$'s in the
Cartan subalgebra.
The action of  an element in $S_{N}$ is given by
\eqn\weyl{X_a^i\to M_a^b X_b^i, \qquad \psi_a^\alpha\to  M_a^b
\psi_b^\alpha,}
where the vector index runs over $i=1,\cdots, 9$  and the spinor
index $\alpha=1,\cdots 16$.

The fermion fields $\psi$ satisfy the following anticommutation
relations,
\eqn\fermanti{ \{ \psi_a^\alpha,\psi_b^\beta\}=
\delta_{ab}\delta^{\alpha\beta}.}
It is convenient to build up the fermionic Hilbert space by
defining fermionic
creation and anihilation operators
$\psih^\alpha_a, {\psih}_a^{\dagger\alpha}$
by
\eqn\creat{\psih^\alpha_a={1\over \sqrt
2}\left(\psi_a^{2\alpha-1}+i\psi_a^{2\alpha}\right),\quad
{\psih}_a^{\dagger\alpha} ={1\over \sqrt
2}\left(\psi_a^{2\alpha-1}-i\psi_a^{2\alpha}\right),\quad
\alpha=1,\cdots ,8,}
which satisfy $\{{\psih}_a^{\dagger\alpha},\psih^\beta_b\}=
\delta_{ab}\delta^{\alpha\beta}$.
A general wave function  $\mid \Psi\rangle$ can be expanded as
\eqn\wavef{ \mid \Psi\rangle = \left(\Psi^{(0)}(X_a)
+\Psi^{(1)}_{a_1}(X_a)\psih^\dagger_{a_1}+{1 \over
2}\Psi^{(2)}_{a_1a_2}(X_a)\psih^\dagger_{a_1}\psih^\dagger_{a_2}+\cdots
\right)\mid 0\rangle,}
where the vector and spinor indices are suppressed and the highest
term has
$8\times N$ fermionic creation operators acting on the Fock space vacuum.

The  expression for the bulk contribution to the Witten index for   
$N$ free
D-particles (with $N$ a prime)  is given
by
\eqn\deficitterm{ I^{(N)}_{0\, bulk} =  \lim_{\beta\to 0} \tr(-1)^F  
e^{-\beta
H}.}
The  trace  is taken over gauge invariant states in the
Hilbert space
and hence one has
to insert a
projector on states which are invariant under the Weyl  permutation  
group,
\eqn\project{{\cal P}={1\over N!}\sum_{\pi\in S_{N}} M_\pi.}
The matrix  $M_\pi$ representing $\pi \in S_N$  act as in \weyl\ on the
coordinates $X_a$
and the
fermions $\psih_a$ in the wave function $\Psi$ given in \wavef.
The bulk index for the free theory  is therefore given by
\eqn\traceb{ I^{(N)}_{0\, bulk} =  \lim_{\beta\to 0}<\Psi| (-1)^F  
e^{-\beta H}
{\cal
P}|\Psi>.}
The factor of  $(-1)^F$ counts bosons (even number of $\psih$) with
$+1$ and fermions (odd number of $\psih$) with $-1$. Because the fermions
transform under the symmetric group matrices $M$, the introduction of the
projector ${\cal P}$ can give a non-vanishing contribution.

The action of $M_\pi$ on   $\psih_a$ and   $X_a$ in
\traceb\
factorizes into a trace over the Hilbert space build from the fermionic
creation
operators and a bosonic gaussian integral coming from the heat kernel
approximation for the free propagator in the limit $\beta\to 0$ \yia.
\eqn\oneterm{I^{(N)}_{0\, bulk}= \lim_{\beta\to 0}{1\over  
N!}\sum_{\pi\in S_N}
\tr_\psi\left( (-1)^F M_\pi\right) \int   
\prod_{i=1}^9\prod_{a=1}^{N-1} dX^i_a
\; {
e^{-(X-M_\pi X)^2/2\beta} \over (2\pi\beta)^{(n-1)9/2}}.}

For all values of $\alpha$  the fermionic trace is  given by
\eqn\fermtracenew{\eqalign {\tr_\psi \left((-1)^F M_\pi\right) &=  
\langle 0\mid
0\rangle -
\sum_a \langle 0\mid \psih_a M_{\pi\,ab}\psih^\dagger_b\mid 0\rangle\cr
&+  {1\over 2}\sum_{a,b}\langle 0\mid  \psih_a   \psih_b
M_{\pi\, ac}M_{\pi\, bd}\psih^\dagger_c\psih^\dagger_d\mid  
0\rangle+\cdots\cr
&= 1- \tr(M_\pi)+{1\over 2} \tr(M_\pi^2)-{1\over 2}  
\tr(M_\pi)^2+\cdots \cr
& =  \det (1-M_\pi), \cr
  }}
so that the trace over all eight fermion components gives a factor of
$\det(1-M_\pi)^8$.     The gaussian integration over the  
coordinates $X^i_a$
similarly gives a factor of $\det(1- M_\pi)^{-9}$

The determinant is easily evaluated by using  an explicit  
representation for
the
  matrices $M_\pi$.  Recall that  these represent the action of  
$S_N$ on the
elements of the Cartan subalgebra.   The roots of $SU(N)$ are the  
vectors in
$R^N$,
\eqn\roots{e_i-e_j,\quad\quad i\neq j;\quad i,j=1,\cdots,N,}
where $e_i$ the $i$th unit vector of $R^N$.
All the roots lie in a
($N-1$)-dimensional subspace $\cal{R}_\omega$  orthogonal to the
vector $\omega=e_1+e_2+\cdots+e_N=(1,1,\cdots,1)$ and an element
$\pi\in S_N$  acts as a permutation, $\pi: e_i\to e_{\pi(i)}$.

Two clases of permutations need to be distinguished:\hfill\break
\noindent (a){\it  Cyclic permutations}\hfill\break\noindent
 These can be
represented   up to conjugation by the  $N\times N$ matrix
\eqn\cycmat{M_{c} = \pmatrix{0  & 1& 0 & 0 & \cdots & 0 \cr
                    0  & 0  &1 & 0 & \cdots & 0 \cr
                    0  & 0  &0 & 1 & \cdots & 0 \cr
                    \cdots & \cdots & \cdots & \cdots & \cdots &
\cdots \cr
            1   & 0  &0  & 0 & \cdots & 0 \cr},}
which has eigenvalues
\eqn\eigencyc{\lambda_k=e^{2\pi i k\over N},\quad k=0,1,\cdots,N-1.}
The matrix  $M_{c}$ does not leave any of the roots  \roots\   
invariant and
 $\omega$ is
the unique eigenvector with eigenvalue  $\lambda_0=1$.   We are  
interested in
evaluating $\det(1-M_\pi) = \det'(1-M_c)$,  where the prime indicates the
omission of the zero eigenvalue.  This is the determinant in the space
orthogonal to $\omega$ and is given by the product of non-zero  
eigenvalues,
\eqn\detcyc{\det^\prime(1-M_c)=\prod_{k=1}^{N-1}(1-\lambda_k)
=2^{N-1}\prod_{k=1}^{N-1}\sin{\pi k\over N}= N. }

\noindent (b) {\it  Non-cyclic permutations}\hfill\break\noindent
These  can be represented by
\eqn\ncycmat{M_{nc}=\pmatrix{M_1 & 0 & 0 & \cdots& 0\cr
                      0      & M_2 & 0 & \cdots& 0 \cr
      \cdots&  \cdots & \cdots & \cdots& \cdots  \cr
0      & 0 & 0 & \cdots &M_p
 \cr}}
where $M_1$, $M_2$, $\cdots$, $M_p$,   represent cyclic
permutations of subsets
of elements and can be written in the form of \cycmat.
The matrix $M_{nc}$ has $p>1$  unit eigenvalues  so  there are  $p-1$
eigenvectors $w_j \in R_\omega$ for which
$(1-M_{nc})w_j= 0$ ($j=1,\cdots,p-1$).  As a result,  some elements  
of the
Cartan subalgebra are left
invariant by $M_\pi$  and
\eqn\nceq{\det(1-M_\pi)= \det'(1-M_{nc}) = 0,}
where the prime again indicates the omission of the zero eigenvalue  
associated
with $\omega$.

To be more precise, a  zero eigenvalue of the bosonic determinant
$\det(1-M_{nc})$   should be interpreted as the inverse volume,   
$R^{-1}$  in
the limit $R\to \infty$, whereas the fermionic determinant vanishes
identically.
  This means that only  the cyclic permutations contribute to the  
index and the
expression \oneterm\ reduces to
\eqn\freeres{I_{0\, bulk}^{(N)} = \sum_{c\in S_N} {1\over
N!}(\det'(1-M_c))^{-1} .}
The non-vanishing determinant  $\det'(1-M_c)=N$ simply counts the winding
number
of a
fundamental  D-particle world-line  and the result  is  the obvious
generalization of the $N=2$ case.
The non-cyclic permutations do not contribute to the index but  have the
interpretation of $p$ disconnected  D-particle
world-lines winding around the euclidean circle.  This is a multi  
D-instanton
configuration and the vanishing of the determinant
is due  to the appearance of extra fermionic zero modes.
Since each of
the $(N-1)!$ cyclic elements of $S_N$ gives the same contribution  to
\freeres\
the result is
\eqn\defresult{I^{(N)}_{0\, bulk}= {1\over N^2} = - I_{def}^{(N)}.}
Combining this with  our earlier ansatz that $I^{(N)}_{bulk} =  
Z_{SU(N)} =1+
1/N^2$
leads to the conclusion that
\eqn\resui{I^{(N)}= I_{bulk}^{(N)} + I_{def}^{(N)} =1,}
 for   prime values of $N$.  This is consistent with  the presence  
of at least
one bound state for every prime value of $N$.

Now consider  the  non-prime charge sector  in which $N =\hat m n$   
with $\hat
m, n >1$.   It is
useful  to recall the interpretation of the D-instanton as a  
wrapped euclidean
D-particle world-line.   The contributions to the instanton measure  
  labelled
by $n$ were  associated with the world-line of a charge-$n$  
D-particle winding
  $\hat m$ times.   In the case when $N$ is prime
the only contribution to the deficit term comes  from $n=1$, $\hat  
m = N$,
which
corresponds to the  winding  of a charge-$\hat m$  D-particle  
world-line.   The
other possibility, $n=N$, $\hat m =1$, is the contribution to the  
Witten index
of the threshold bound state.
 It is therefore very compelling to {\it assume} that every  
threshold bound
state  of charge $n$ contributes to the deficit term in the same  
way as the
fundamental ($n=1$) D-particle.   This  is a generalization of the   
 property
proved in the $N=2$ case by \sethia\ that the deficit term is  
determined by
free particle dynamics.
It  means that  the space of free-particle states that enters in the
calculation of of $I_{0\, bulk}$ should be enlarged to include the  
infinite
tower of charge-$n$ threshold bound states.  In that case a cyclic  
permutation
of $\hat m$ identical charge-$n$ particles  contributes  a term
\eqn\nfree{ I^{(\hat m)}_{0\, bulk}= {1\over {\hat m}^2},}
which  leads to a total deficit in the charge-$N$ sector of
\eqn\nmtot{ I^{(N)}_{def} = -\sum_{N|\hat{m}\atop \hat{m}>1}  
I^{(\hat{m})}_{0\,
bulk } = -\sum_{N|\hat{m}\atop \hat{m}>1}{1\over
{\hat{m}}^2} ,}
 consistent with the Witten index $I^{(N)}=1$ for all integer $N$.

The success of these assumptions points to the systematics that  
needs to emerge
from a more precise treatment of the integration over the boundary  
term when
$N$
is not prime.  This suggests that there are regions of  moduli
space in which the non-abelian integration leads to  $\hat m$ charge-$n$
threshold bound states which behave as free particles.

\newsec{Conclusions}
This paper has made   a connection between the measure that  enters
the charge-$N$ D-instanton
corrections to the IIB effective action and the bulk contribution
to the Witten
index  of D-particle quantum mechanics which is related to  the  
existence of
threshold charge-$N$  bound states of D-particles.   The tight  
constraints on
the D-instanton action  implied by the $SL(2,Z)$   duality of
IIB string theory lead to a  conjectured expression for the measure  
in the sum
over single charge-$N$ D-instantons.   This should encode the
value of the  the
complicated
$SU(N)$ integrals that need to be evaluated if the measure is calculated
directly from \zdef. Related issues arise in the considerations of  
\bachas\
concerning D-string instantons in $d=8$ type I string theories.

 This is reminiscent of  the Seiberg-Witten  arguments concerning the
solution of $N=2$
supersymmetric Yang--Mills \seibergwit.  In that case duality  
constrained  the
form  of the
measure  of the instanton corrections to the prepotential without  
explicit
evaluation of the complicated
integrals over the ADHM-moduli space \dorey.  In this regard it is also
interesting  to note
that the
connection between the non-perturbative D-instanton corrections to
the effective
action in ten dimensions  and a regularized one loop term in eleven
dimensional
M-theory  \ggv\ also has an analogy
in $N=2$ supersymmetric Yang--Mills,
where the instanton-corrected prepotential
 can be related to a one-loop term in
five-dimensional supersymmetric Yang--Mills theory
\nekrasov.
\vskip .2cm

\noindent {\bf Acknowledgements}

We are  grateful to S. Sethi and W. Taylor for useful
conversations. The work
of M. Gutperle was partially supported by DOE grant
DE-FG02-91ER40671, NSF
grant
PHY-9157482 and a James. S. McDonnell grant 91-48.
\vskip .2cm

\listrefs

\end